\documentclass[aps,prl,showpacs,twocolumn,preprintnumbers,amsmath,groupedaddress]{revtex4}
\usepackage{bm}
\usepackage{amsmath}
\usepackage[dvips]{graphicx}
\newcommand{\vF}{\ensuremath{\nu_\text{F\,}}}
\newcommand{\bsigma}{\ensuremath{\bm\sigma}}
\newcommand{\bp}{\ensuremath{\bm {p}}}
\newcommand{\br}{\ensuremath{\bm {r}}}

\newcommand{\Ncal}{\mathcal{N}}

\newcommand{\e}{\epsilon}
\newcommand{\gt}{\tilde{g}}

\DeclareMathOperator{\sign}{sign}
\DeclareMathOperator{\real}{Re}

\begin{document}


\title{Adatoms in graphene}


\author{A.~H. Castro Neto}
\affiliation{Department of Physics, Boston University, 590 
Commonwealth Avenue, Boston, MA 02215, USA}

\author{Valeri N. Kotov}
\affiliation{Department of Physics, Boston University, 590 
Commonwealth Avenue, Boston, MA 02215, USA}

\author{Johan Nilsson}
\affiliation{Instituut-Lorentz, Universiteit Leiden, P.O. Box 9506, 2300 RA Leiden, The Netherlands}

\author{Vitor M. Pereira}
\affiliation{Department of Physics, Boston University, 590 
Commonwealth Avenue, Boston, MA 02215, USA}

\author{N. M. R. Peres}
\affiliation{Centro de F\'isica e Departamento de F\'isica, Universidade do Minho, P-4710-057, Braga, Portugal}

\author{Bruno Uchoa}
\affiliation{Department of Physics, University of Illinois at Urbana-Champaign, 1110 W. Green St., Urbana, IL 61801-3080}

\date{\today}


\begin{abstract}
We review the problem of adatoms in graphene under two complementary points of view, scattering theory and strong correlations. We show that in both cases impurity atoms on the graphene surface present effects that are absent in the physics of impurities in ordinary metals. We discuss how to observe these unusual effects with standard experimental probes such as scanning tunneling microscopes, and spin susceptibility. 
\end{abstract}

\maketitle


It has been known since the discovery of graphene \cite{novo,Geim_review}
that disorder plays a fundamental role in determining its physical
properties \cite{nuno}. It was understood early on that the linear 
dependence of the graphene conductivity with electronic 
density, or gate bias, at large densities 
requires the presence of long range scattering, either in the form of
Coulomb interactions \cite{nomura_2,nomura,adam}, 
or correlated ripples \cite{kats_ripples}. Experiments with K atoms
in ultra high vacuum conditions have shown that the conductivity is
affected by ionized impurities \cite{fuher}, although that might
depend strongly on the screening properties of the environment \cite{mohi}.
In fact, because of its electrical sensitivity to adatoms, it is possible to 
use graphene as a single molecule detector \cite{schedin}.

Although unwanted adatoms can be a nuisance for transport properties, one
can also think of them as a way to tailor graphene, in order to create
new many-body states that do not show up in pure
graphene \cite{uchoa_tailor}. The situation here is akin to what is found
in ordinary metals which, when in pure form, are weakly interacting and
inert, but show extraordinary properties such as magnetism,
Kondo effect and superconductivity, when doped with transition metal atoms 
with strongly interacting inner electronic shells \cite{hewson}. 

The advantage of graphene in comparison to ordinary three dimensional (3D)
metals (where adatoms are introduced by alloying, which is generically
a random process) is that graphene is an open surface, and hence adatoms 
can be manipulated with the use of atomic force microscopy to obtain 
structures with atomic precision \cite{eigler}. Only recently, 
has scanning tunneling microscopy
(STM) been used to study adatoms in graphene \cite{crommie}, and still
very little is known about the spectroscopic properties of these adatoms.

Historically there have been two main approaches for the understanding of
impurity atoms in ordinary metals. The most traditional one, championed
by J. Friedel, follows the arguments of traditional scattering theory
in terms of the scattering phase shifts of the electrons in the
presence of an external potential \cite{friedel}. In this kind of 
approach, the electrons of the impurity atom and the metal host 
are treated at equal footing and the many-body aspects of the problem
enter only through Pauli's exclusion. As shown by Anderson \cite{anderson},
this kind of approach is inadequate for atoms with inner shell
electrons, where the Coulomb interaction is large and has to
be taken into account from the outset. In this case, one distinguishes
between the host and the adatom electrons. These two approaches are 
essentially complementary and serve to address different issues in the
problem of impurity atoms in metallic hosts.

We will discuss these two approaches in the context
of adatoms in graphene. In the fist part of this paper we discuss the 
Friedel approach. We are going to show
that the Coulomb problem in graphene has features similar to the ones
found in the 3D relativistic Coulomb problem. In particular, the
famous ``fall to the center effect'' \cite{Landau-QM:1981} 
that would occur in atoms with
atomic number $Z$ greater than $1/\alpha$, where 
$\alpha = e^2/(\hbar c)$ is the fine structure constant, 
can occur in graphene even in the presence of a single proton
if the dielectric constant $\varepsilon_0$ is of the order of $1$. This effect
can be observed by STM spectroscopy by studying the local density
of states (LDOS) close to an impurity. It is interesting that graphene
allows for an opportunity to study an effect that was predicted to occur
only in heavy ion collisions, or in the event horizon of a black hole
\cite{greiner_muller}. 

In the second section, we present the generalization
of the Anderson impurity model to graphene, taking into account the
local Coulomb repulsion (Hubbard term) of inner shells. We show that
the Anderson problem in graphene has features that are rather different from
the ones found in ordinary metals \cite{anderson}. In particular, we argue
that transition metals that would not be magnetic in ordinary metals can
become magnetic in graphene due to its peculiar properties. In the last
section we include our conclusions.

\section{Friedel's way \cite{vitor_coulomb,shytov:2007,novikov:2007,falko}}     

In what follows, we use the effective description
of the electrons at low energies and long wavelengths \cite{neto_review} 
by expanding the electronic dispersion 
around the Dirac point in the Brillouin zone 
(K and K' points) and obtaining the two-dimensional Dirac (we use
units such that $\hbar=1$):
\begin{eqnarray}
H_0 = \vF \hat{\bsigma} \cdot \bp
\label{dirac}
\end{eqnarray}
where $\vF \approx 6$ eV \AA \, \, is the Fermi-Dirac velocity \cite{neto_review}, 
$\sigma_i$ ($i=1,2,3$) are Pauli matrices, and $p_i$ ($i=x,y$) is the 
two-dimensional momentum operator. Consider the problem of a single Coulomb 
impurity with charge $Z e$ sitting at the origin. 
The wavefunction of the problem obeys the equation:
\begin{equation}
  \vF \Bigl(-i 
    \hat{\bsigma}\cdot \nabla - \frac{g}{r}  
  \Bigr)
  \Psi(\br) = E\;\Psi(\br)
  \label{eq:SchrodingerEq}
  ,
\end{equation}
where $\Psi(\br) = (\psi_A(\br),\psi_B(\br))$ is a spinor wavefunction, and
\begin{eqnarray}
 g \equiv \frac{Z e^2}{\vF\varepsilon_0} \, ,
\label{fine}
\end{eqnarray}
is the dimensionless coupling constant of the problem. Hence, the coupling constant depends directly on the dielectric properties of the medium where graphene is sitting \cite{jang_ice}.

Notice that Eq.(\ref{eq:SchrodingerEq}) only involves one
of the Dirac cones. This assumption is justified on the basis that the Coulomb interaction is long-ranged, and hence singular in momentum space (it behaves as $1/p$ as $p \to 0$). Therefore, scattering is strongly localized around one of the cones. We have checked this assumption numerically by solving the Coulomb problem in the tight-binding model in the lattice \cite{vitor_coulomb}. 

Observe that the total angular momentum, $J_z=L_z + \sigma_z /2$ \cite{DiVincenzo:1984}, is conserved and we can therefore diagonalize the problem in the basis of momentum states:
\begin{equation}
  \psi_j(\br) = \frac{1}{\sqrt{r}} 
  \binom{e^{i(j-\frac{1}{2})\varphi} \varphi^A_j(r)}
  {i e^{i(j+\frac{1}{2})\varphi} \varphi^B_j(r)}
  , 
  \label{eq:Def-Spinor}
\end{equation}
in cylindrical coordinates, Eq. \eqref{eq:SchrodingerEq} becomes:
$(j=\pm1/2,\,\pm3/2,\,\ldots)$
\begin{equation}
  \begin{pmatrix}
    \e +g/r &  -(\partial_r +j/r) \\
    (\partial_r - j/r) & \e + g/r
  \end{pmatrix}
  \binom{\varphi^A_j(r)}
  {\varphi^B_j(r)} = 0
  .
  \label{eq:SchrodingerEq-Radial}
\end{equation}
The solutions for this problem can be written as:
\begin{equation}
  \varphi_j(r) = \sum_{n =\pm 1} C_n u_n f_n(r) 
  \,,\,
  u_{\pm 1} = \sqrt{\frac{1}{2|j|}} 
    \binom{\sqrt{|j\pm\lambda|}}{s_{gj} \sqrt{|j\mp\lambda|}}
  ,
  \label{eq:DiagonalWF}
\end{equation}
where $s_x \equiv \sign(x)$,
\begin{eqnarray}
\lambda=\sqrt{j^2-g^2} \, ,
\label{lamb}
\end{eqnarray}
 and $f_\lambda(r)$ is the solution of the equation:
\begin{equation}
  \partial_r^2 f_{\pm 1}(r) + 
  \left[
    \e^2 + \frac{2g\e}{r} - \frac{\lambda(\lambda \mp 1)}{r^2}
  \right]
  f_{\pm 1}(r) = 0 \, ,
  \label{eq:CoulombEq}
\end{equation}
which is the radial equation for the
3D Coulomb problem \cite{Landau-QM:1981} 
with the energy, $E$, replaced by $\e^2$,
and the angular momentum quantum number, $\ell$, replaced by $\lambda$.

We immediately notice from (\ref{lamb}) that $\lambda$ can be a purely
imaginary number when $g > j$. The presence
of ``imaginary angular momentum'' channels in the problem can be
initially troublesome but they are mathematically allowed. Notice
that, since $j$ is a half-integer, 
the smallest value of $g$ for this to happen occurs for $g=g_c=1/2$.

\begin{figure}[htb]
\begin{center}
\includegraphics*[clip,width=1.0\columnwidth]{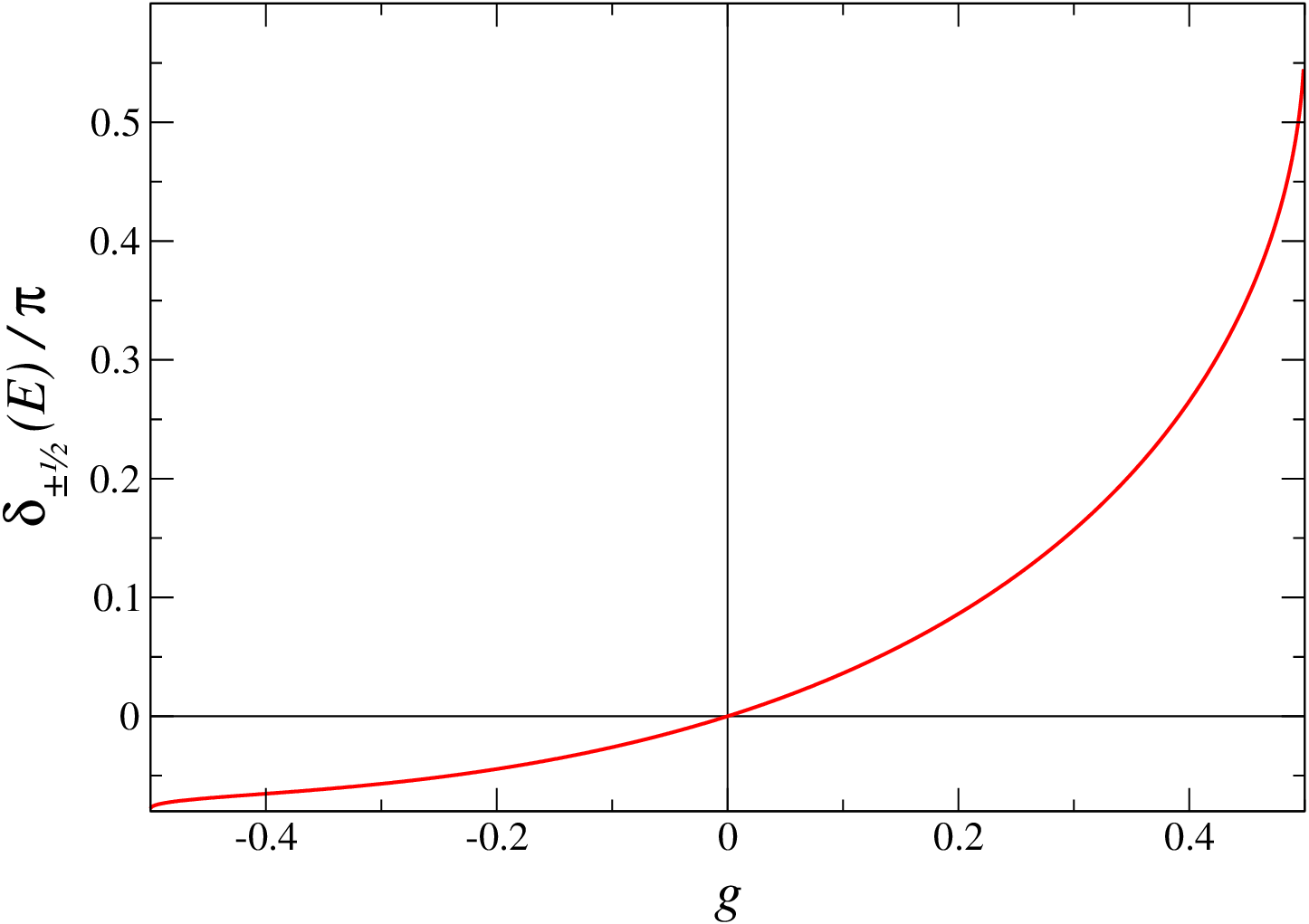}
\end{center}
\caption{(color online) Phase shift of the Coulomb problem in the
undercritical regime as a function of coupling $g$.}
\label{phase_under}
\end{figure}

For $g<g_c$, the so-called undercritical regime, 
eq.~\eqref{eq:CoulombEq} can be solved as (see \eqref{eq:DiagonalWF}):
\begin{equation}
  \varphi_j(r) / \Ncal_j = 
    u_+ F_{\lambda-1}(-\gt,\rho) + s_{g\e} u_- F_{\lambda}(-\gt,\rho)
  ,
  \label{eq:RadialSolution}
\end{equation}
where $\Ncal_j$ is a normalization factor, $F_L(\eta,\rho)$ is a Coulomb
wave functions \cite{Abramowitz:1964,Yost:1936}, $\rho = |\epsilon| r$, 
and $\gt=s_\e g$. At long distances, the asymptotic behavior of this
function is given by \cite{Landau-QM:1981}:
\begin{eqnarray}
  F_\lambda(-\gt,\rho) \sim \sin\Bigl(
    \rho + \gt \log(2\rho) + \vartheta_\lambda(\gt)
  \Bigr), 
  \label{eq:Coulomb-Phase-Shift}  
\end{eqnarray}
where $\vartheta_\lambda(\gt) = - \lambda \frac{\pi}{2} +
    \arg\bigl[\Gamma(1+\lambda-i\gt)\bigr]$.
Hence, in 
\eqref{eq:RadialSolution} we have asymptotically: 
\begin{equation}
 \varphi_j(r) \sim  
\sin\Biggl[ 
    \rho + \gt \log(2\rho) +
    \arg\Bigl(
      u_+ e^{i\vartheta_{\lambda-1}}+s_{g\e} u_- e^{i\vartheta_{\lambda}}
    \Bigr)
  \Biggr]
  .
  \label{eq:Our-Phase-Shift}  
\end{equation}
One notices the logarithmic phase shift which is traditional in Coulomb problems. Observe that the phase shift is energy {\it independent} in this case, but it is strongly dependent on the value of the coupling $g$. This result has implications to the so-called Friedel sum rule. In Fig.\ref{phase_under} we plot the value of the phase shift as a function $g$.

The presence of the adatom modifies the local density
of states (LDOS):
\begin{eqnarray}
N(\epsilon,r) &=&
\sum_E |\Psi_E(r)|^2 \delta(\epsilon-E) 
\nonumber
\\
&=& \sum_{j=-\infty}^{\infty} n_j(\e,r) \, ,
\end{eqnarray}
where
\begin{eqnarray}
n_j(\e,r) &\equiv& r^{-1} |\varphi_j^A(r)|^2 + r^{-1} |\varphi_j^B(r)|^2
\nonumber
\\ 
&=& (\Ncal_j^2/r)
  \left[
    F_{\lambda-1}^2 + F_{\lambda}^2 + 2\gt F_{\lambda}F_{\lambda-1}/|j|
  \right]
  .
  \label{eq:LDOS-Weak}
\end{eqnarray}
is the contribution of each momentum channel to the LDOS. 
In the limit $\gt\to0$ the Coulomb wave functions reduce to Bessel functions
\cite{Abramowitz:1964}, and one obtains the free DOS \cite{neto_review} 
if the normalization is chosen as $\Ncal_j^{-2}=2\pi^2\lambda^2/j^2$.
It is clear from \eqref{eq:LDOS-Weak} that
particle-hole symmetry is broken, and as $\e \to 0$ we have
\begin{eqnarray}
N(\e,\br)\sim |\e|^{2\lambda} \, ,
\label{neb}
\end{eqnarray} 
and $\lambda<1/2$ for $|j|=1/2$. At higher energies, or far away from the adatom, we obtain the bare density of states of graphene, namely, $N_{0}(\e,\br) \sim |\e|$.  

\begin{figure}[t]
\begin{center}
\includegraphics*[clip,width=1.0\columnwidth]{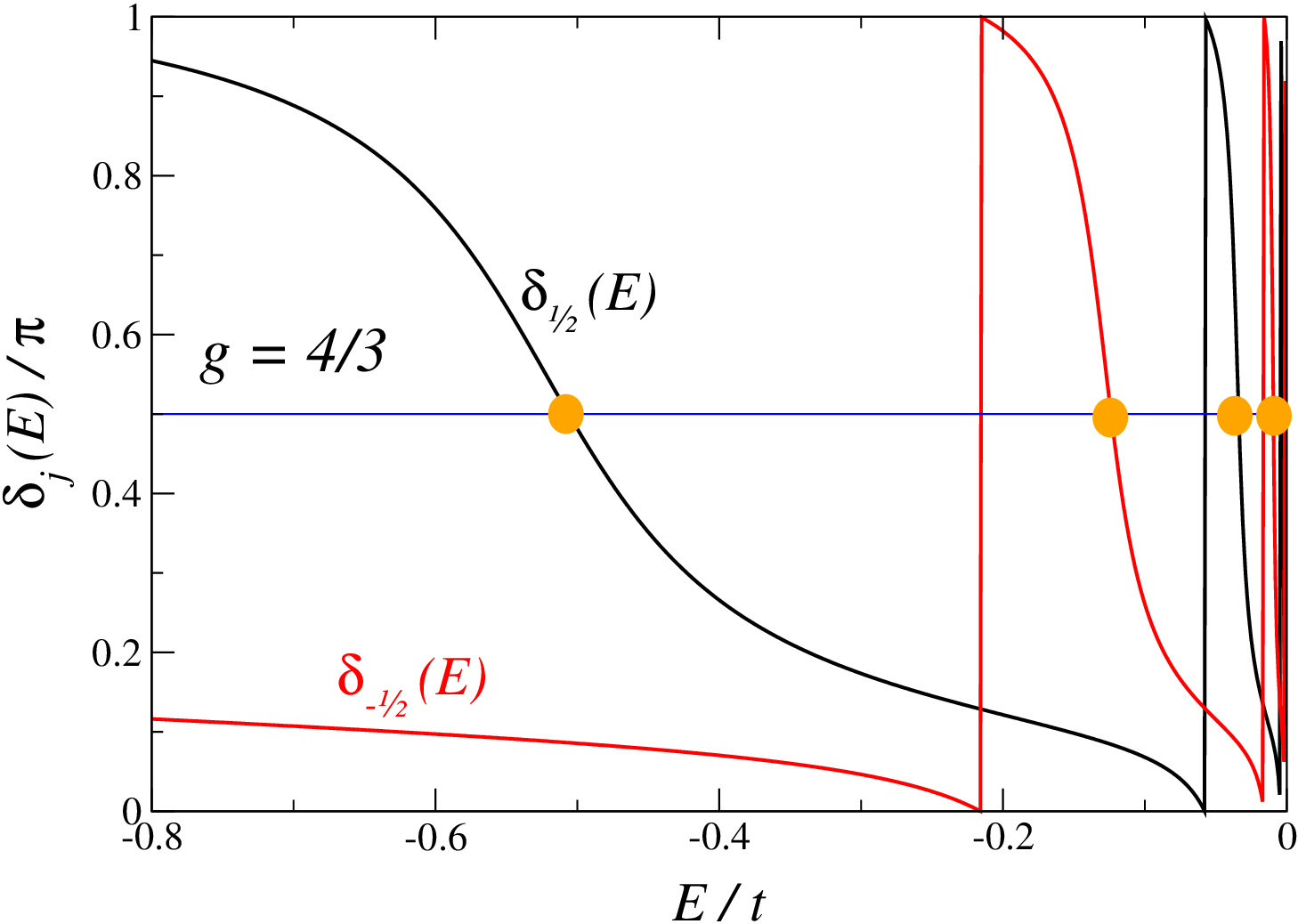}
\end{center}
\caption{(color online) Phase shift of the Coulomb problem as a function of energy $E$ in the supercritical regime ($g>g_c$).}
\label{phase_over}
\end{figure}

For $g>g_c$, the so-called supercritical regime, 
the solution of \eqref{eq:RadialSolution} is of the form, 
$
\bar{\varphi}_j(r) = C_1 \; \bar{\varphi}_{i\beta}(r) + C_2 \; \bar{\varphi}_{-i\beta}(r) \, ,
$
where $C_{1,2}$ are to be set by the boundary conditions at short distances and
\begin{equation}
  \bar{\varphi}_{i\beta}(r)  = 
    \bar{u}_+ F_{i\beta-1}(-\gt,\rho) + 
    s_{jg\e} \bar{u}_- F_{i\beta}(-\gt,\rho)
  ,
  \label{eq:RadialSolution-2}
\end{equation}
$\beta=-i\lambda = \sqrt{g^2-j^2}$ for those $j$'s such that
$|g|>|j|$, and
\begin{equation}
  \bar{u}_{\pm 1} = \sqrt{\frac{1}{2|g|}} 
  \binom{\sqrt{j\pm i\beta}}{s_{g} \sqrt{j\mp i\beta}}
  .
  \label{eq:DiagonalWF-2}
\end{equation}

The issue of boundary conditions at short distances is quite interesting in this case. Notice that as $r \to 0$, the behavior of the Coulomb wavefunctions with imaginary index is rather peculiar:
\begin{eqnarray}
F_{i\beta-1}(-gt,\rho \to 0) \sim \rho^{i\beta} \sim \cos(\beta \ln(\rho)) \, ,
\end{eqnarray}
and, hence, the wavefunction oscillates endlessly, with an infinite number of zeros, as one approaches the origin. This is a rather puzzlingly feature since in basic quantum mechanics the energy of a state is directly related to the number of zeros of the wavefunction. 

This effect is known as the ``fall of a particle to the center''\cite{Landau-QM:1981} and occurs when the external potential is too strong \cite{greiner_muller}. Notice that the condition $g>g_c = 1/2$ is equivalent, in terms of (\ref{fine}), to:
\begin{eqnarray}
Z > \frac{1}{2 \alpha_G}
\end{eqnarray}
where 
\begin{eqnarray}
\alpha_G = \frac{e^2}{\vF \varepsilon_0} \, ,
\end{eqnarray}
 is graphene's ``fine structure constant'' \cite{note,kotov}. In relativistic quantum mechanics this effect shows up as the ``fall'' of the electron into the nucleus when the atomic number is too high. As the electron falls, a positron is emitted leading to the charging of the vacuum. The analogy in graphene is clear: if the potential is attractive to the electrons, the electrons become strongly bound to the adatom and a ``hole'' is ejected. Unlike the problem in free space, the problem at hand occurs in a lattice which always has a natural ultraviolet cut-off, namely, the lattice spacing or in this case the distance of the adatom from the carbon atoms, $a_0$. Hence, this indefinite oscillation of the wavefunction is controlled by a short distance cut-off $a_0$. The vanishing of the wavefunction at $r<a_0$ translates \cite{Berry:1987} into the boundary condition: $\varphi_j^A(a_0)=\varphi_j^B(a_0)$. This allows us to fix the value of $C_1/C_2$ and determine the total scattering phase shift in the supercritical regime.
Unlike the undercritical case, the phase shift is a strong function of the energy, as shown in Fig.~\ref{phase_over}.

\begin{figure}[t]
\begin{center}
(a) \includegraphics*[clip,width=0.4\columnwidth]{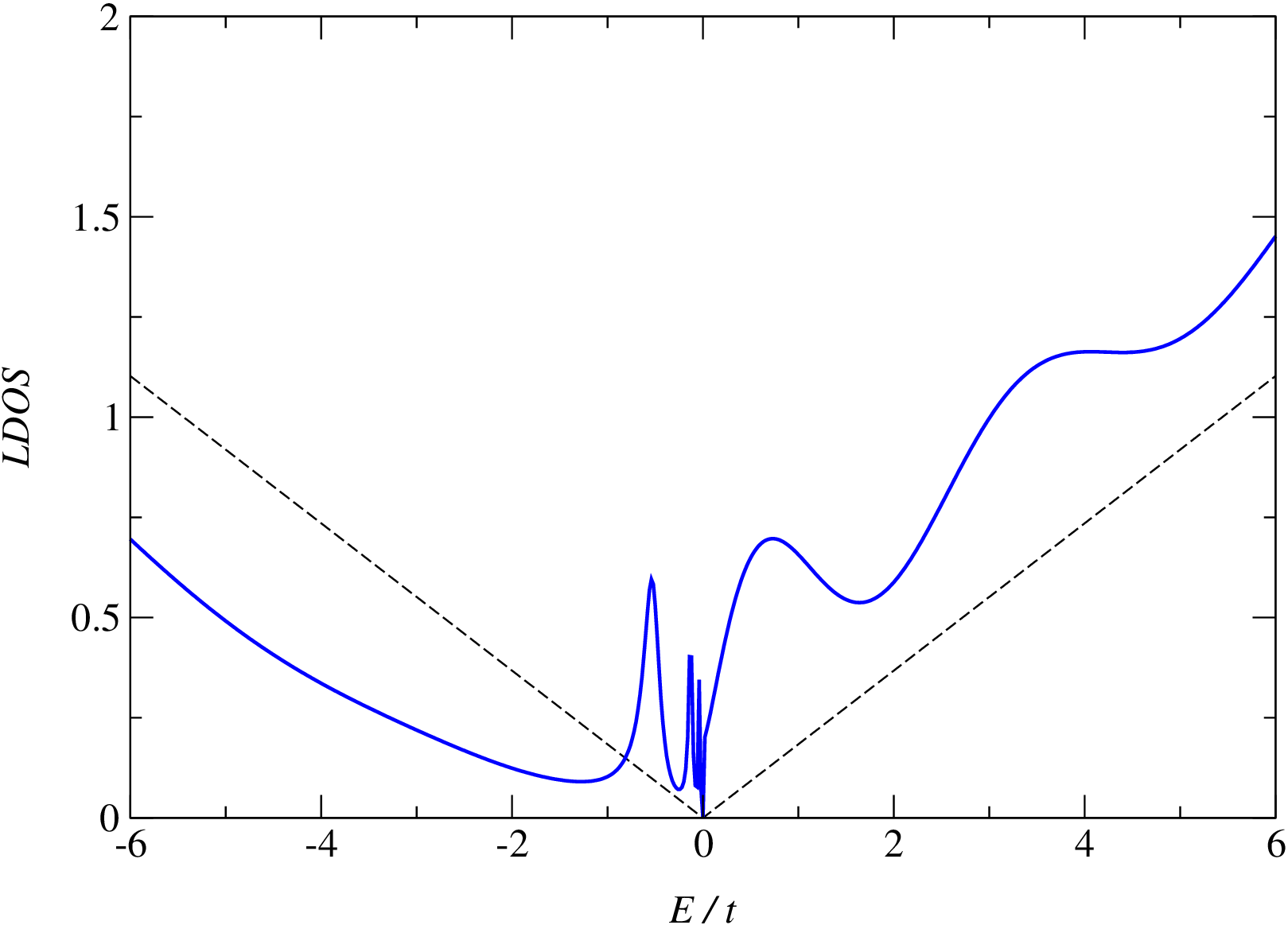}
(b) \includegraphics*[clip,width=0.4\columnwidth]{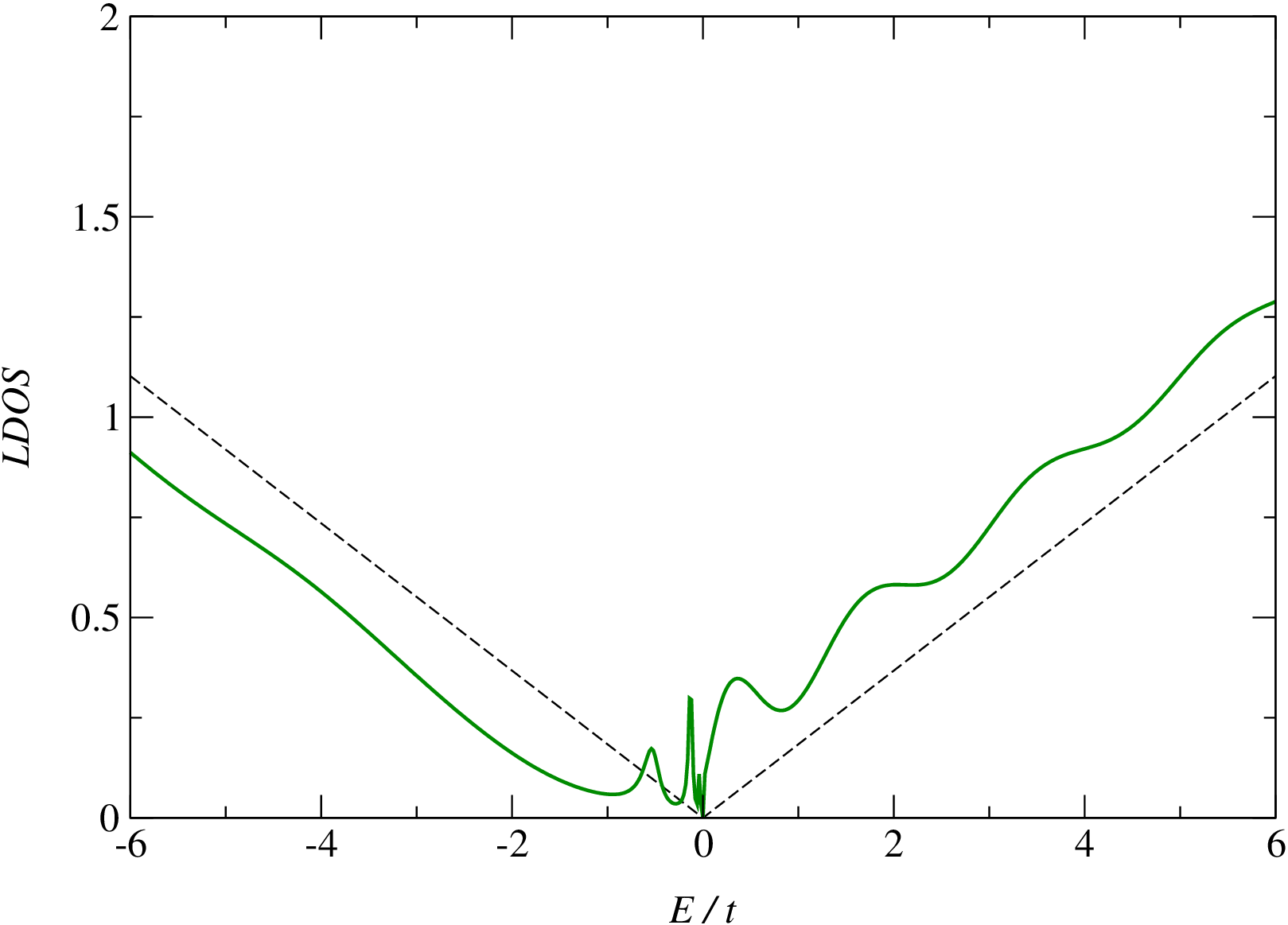}
\end{center}
\begin{center}
(c) \includegraphics*[clip,width=0.4\columnwidth]{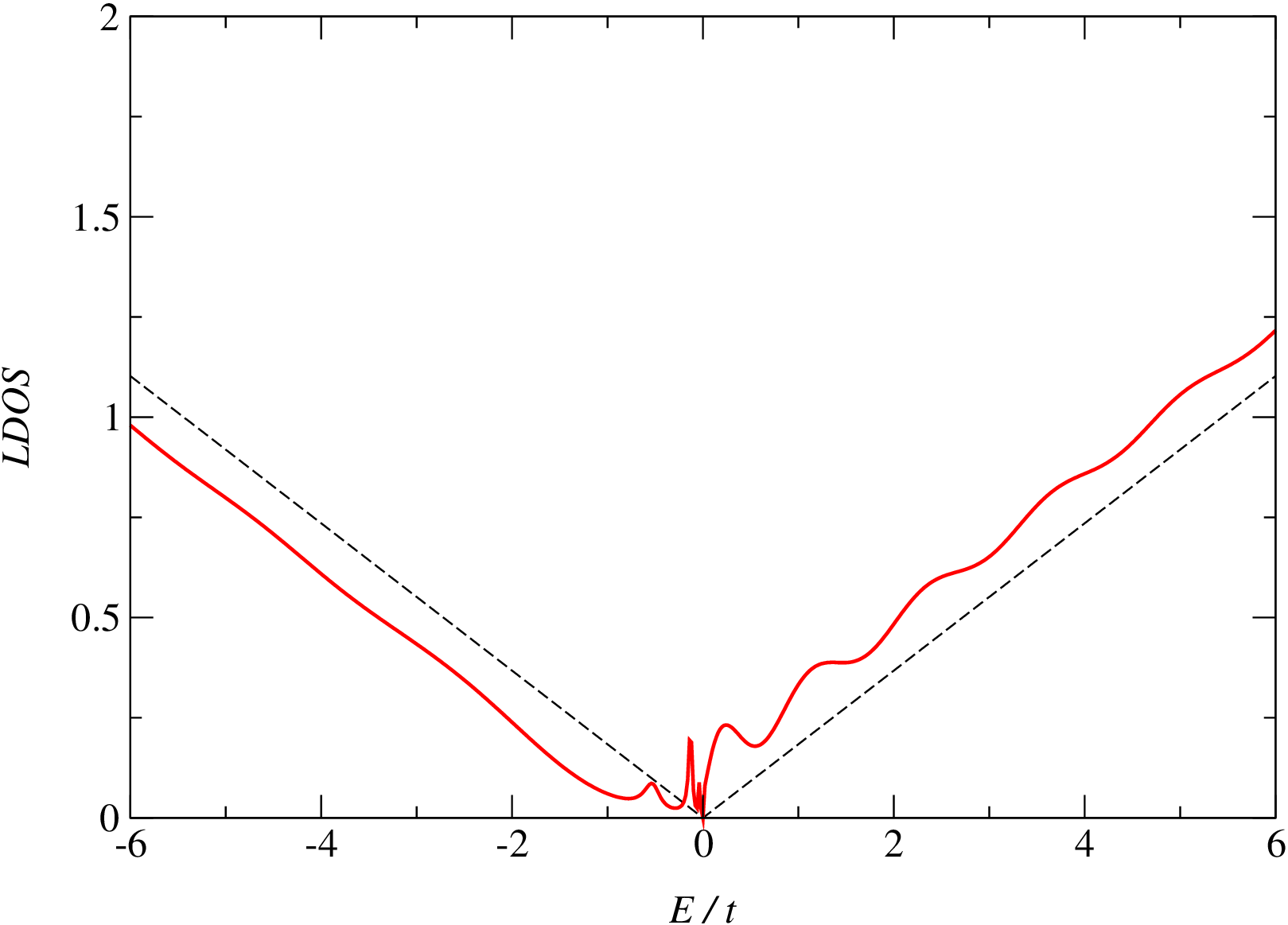}
(d) \includegraphics*[clip,width=0.4\columnwidth]{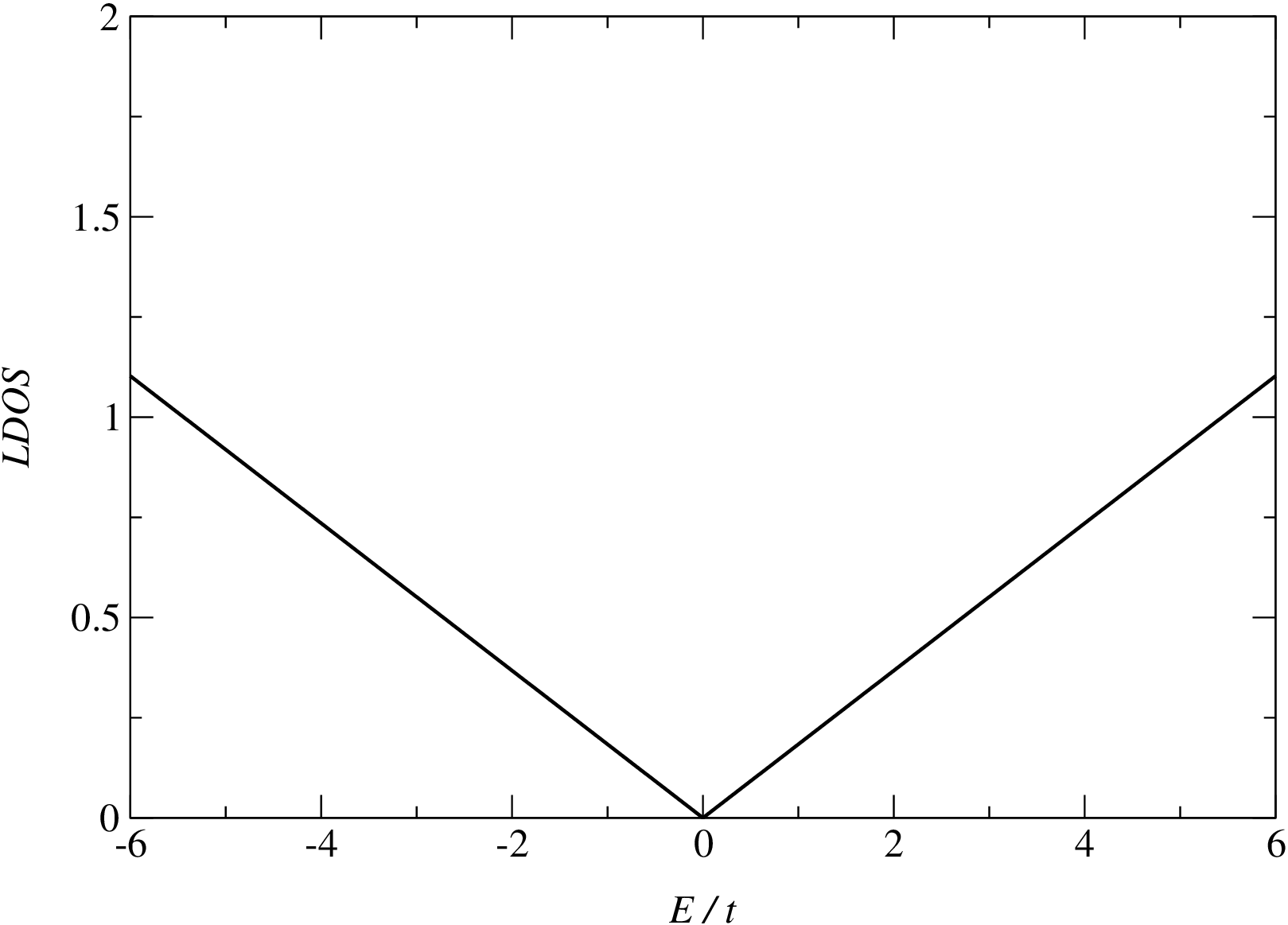}
\end{center}
\caption{(color online)LDOS as a function of energy at different
distances $r$ from the adatom: (a) $r=a_0$; (b) $r=2 a_0$; (c) $r=3 a_0$; (d) $r \to \infty$.}
\label{LDOS_over}
\end{figure}

We can now proceed and compute the contribution of the overcritical $j$'s 
to the LDOS:
\begin{subequations}
  \label{eq:LDOS-Strong}  
\begin{equation}
    \bar{n}_j(\e,r) = \frac{1}{2 \pi^2 r}
    \frac{
      \varrho_j^I(\rho) + 
      s_{\e j} \real\bigl[e^{i 2 \delta_j } \varrho_j^{II}(\rho) \bigr]
    }{
      \varrho_j^I(\infty) + 
      s_{\e j} \real\bigl[e^{i 2 \delta_j } \varrho_j^{II}(\infty) \bigr]
    }
    ,
\end{equation}
where,  
\begin{align}
    \varrho_j^I &\equiv
    |F_{i\beta }|^2  + |F_{i\beta -1}|^2 + 
    2 |j| \real[F_{i\beta} F_{-i\beta -1}]/\gt
    ,\\
    \varrho_j^{II} &\equiv 
    2 F_{i\beta} F_{i\beta -1} + |j| 
    (F_{i\beta }^2 + F_{i\beta -1}^2)/\gt
    .
\end{align}
\end{subequations}
Eqs.~\eqref{eq:LDOS-Weak} and \eqref{eq:LDOS-Strong} determine the LDOS for any
coupling strength, $g$, which can be summarized as 
\begin{equation}
  N(\e, \br) = \sum_{|j|<|g|} \bar{n}_j(\e,r) + \sum_{|j|>|g|} n_j(\e,r)
  .
  \label{eq:LDOS-Any}
\end{equation}
In Fig.\ref{LDOS_over} we show the LDOS as a function of energy at different distances from the adatom. One can clearly see formation of resonances in the LDOS that are directly associated with the positron creation of the actual relativistic problem. The possibility of testing ideas of relativistic physics in a carbon-based material is unique to graphene due to the robustness of its Dirac spectrum.

In the previous discussion we have completely disregarded the Coulomb interaction between the electrons. There are three types of electron-electron interactions one should worry about: electron-electron interactions within graphene electrons, interactions between graphene electrons and adatom electrons, and between electrons in the adatom.  Electron-electron interactions within graphene, and between electrons in graphene and in the adatom, can change some of the quantitative features discussed here, as we shall see below. The electron-electron interactions within the adatom can lead to new physics, namely, the possibility of creating a local magnetic moment in the adatom. This kind of situation is not contemplated in Friedel's way and will be the subject of next section.

Let us consider the problem of screening in graphene. Within the random phase approximation (RPA), in the static case, 
the screened potential of an external impurity is given by \cite{fetter}:
\begin{eqnarray}
V_s(q) = \frac{V_0(q)}{1-\Pi^{(0)}(q) U(q)}
\label{effpot}
\end{eqnarray}
where $V_0(q) = Z U(q)$ is the Fourier transform of the impurity Coulomb potential in 2D, $U(q) = 2 \pi e^2/q$ the Fourier transform of the bare electron-electron interaction, and $\Pi^{(0)}(q)$ is the static polarization function. In the above expression we have disregarded the modification of the dielectric function due to electron-electron interactions \cite{kotov_die}.
In a metal, we have $\Pi^{(0)}(q \to 0) \propto - N(E_F)$, the density of states in the Fermi level, and hence $V_s(q \to 0) \to $ constant, indicating that the potential is completely screened. 

Another way to understand screening is by computing the total induced charge density. The local particle density in RPA is given by \cite{fetter}:
\begin{eqnarray}
\delta n({\bf r}) = - \int d{\bf q} 
\frac{\Pi^{(0)}(q) V_0(q)}{1-\Pi^{(0)}(q) U(q)} e^{i {\bf q} \cdot {\bf r}}
\end{eqnarray}
and hence the total induced particle density is:
\begin{eqnarray}
\delta N &=& \int d{\bf r} \delta n({\bf r})
 \nonumber
\\
&=& - \frac{\Pi^{(0)}(q \to 0) V_0(q \to 0)}{1-\Pi^{(0)}(q \to 0) U(q \to 0)}
\label{dQ}
\end{eqnarray}
and one immediately sees, in very general terms, that if $V_0(q \to 0)$ 
and $U(q \to 0)$ are divergent and $\Pi^{(0)}(q \to 0)$ is regular, as in the case of a normal metal, the RPA result is that 
\begin{eqnarray}
\delta N_{{\rm metal}} = Z \, ,
\label{nmetal}
\end{eqnarray}
that is, the local impurity is completely screened.

Let us apply the same reasoning to graphene. Because RPA is valid in linear response, it only applies to weak coupling, that is, in the undercritical regime, $g<g_c$. The polarization function
in graphene has been calculated many times before \cite{s86,WSSG06,HdS06},  
but the argument above only requires knowledge of the polarization function at long wavelengths. The asymptotic behavior of the polarization function is:
\begin{eqnarray}
\Pi^{(0)}(q \to 0,\mu=0) \approx - \frac{q}{4 \vF} \, ,
\label{p0}
\end{eqnarray}
for $\mu =0$ (that is, when the Fermi energy is at the Dirac point), and 
\begin{eqnarray}
\Pi^{(0)}(q \to 0,\mu \neq 0) \approx - \frac{2 |\mu|}{\pi \vF} \, ,
\label{p1}
\end{eqnarray}
when $\mu \neq 0$. Replacing these results in (\ref{effpot}) and (\ref{dQ}) give:
\begin{eqnarray}
V_s(q \to 0,\mu=0) &\approx& \frac{Z}{1+\frac{\pi}{2} \alpha_G} \frac{2 \pi e^2}{q} \, ,
\nonumber
\\ 
\delta N(\mu=0) &=& Z \frac{\frac{\pi}{2} \alpha_G}{1 + \frac{\pi}{2} \alpha_G}
\, ,
\label{nm0}  
\end{eqnarray}
for $\mu=0$, and
\begin{eqnarray}
V_s(q,\mu) &\approx& \frac{2 \pi Z e^2}{q+q_0(\mu)} \, ,
\nonumber
\\
\delta N (\mu)  &=&  Z \, ,
\label{nmf}
\end{eqnarray}
for $\mu \neq 0$, where $q_0 = 4 \alpha_G |\mu|$ is the Thomas-Fermi screening
length. This result indicates that graphene screens like a metal
whenever $\mu$ is finite but when the Fermi energy is at the Dirac point
there is a residual unscreened charge, 
\begin{eqnarray}
\delta N_{{\rm uns.}} &=& Z-\delta N(\mu=0) = Z^* \, ,
\nonumber
\\
Z^* &=& \frac{Z}{1+\frac{\pi}{2} \alpha_G} \, ,
\end{eqnarray}
in accordance with (\ref{nm0}) showing the Coulomb potential remains the same even with screening and that the only modification is the value of the local charge that changes from $Z$ to $Z^*$. Notice that the limits of $\mu \to 0$ and $q \to 0$ do not commute in graphene. In fact, the limit of $\mu=0$ and $q \to 0$ is rather pathological.

In the overcritical regime the problem of electron-electron interactions becomes of fundamental importance. As we have seen previously, the ``fall of the particle to the center'' indicates that large amounts of charge accumulate in the vicinity of the impurity. This accumulation of charge implies strong electron-electron interactions beyond the description of RPA. Instead, the screening becomes non-linear \cite{Kscreening,fogler} and the Coulomb law is modified. This regime, however, is highly non-perturbative in nature and it is not clear that the approximations that are usually used in this regime, such as the non-linear Thomas-Fermi theory, are really applicable.      

\section{Anderson's way \cite{uchoa_anderson}}

In Anderson's approach we distinguish between the graphene electrons,
described by Hamiltonian (\ref{dirac}), and the electrons localized
in the inner shell of the adatom which have an Hamiltonian:
\begin{eqnarray}
H_{f}=\epsilon_{0}\sum_{\sigma}f_{\sigma}^{\dagger}f_{\sigma} + 
Uf_{\uparrow}^{\dagger}f_{\uparrow}f_{\downarrow}^{\dagger}f_{\downarrow}
\label{anderson}
\end{eqnarray}
where $f_{\sigma}$ ($f^{\dagger}_{\sigma}$) annihilates (creates) and
electron with spin $\sigma = \uparrow,\downarrow$, 
$\epsilon_{0}$ is the bare energy of the local level, 
and $U$ is the energy cost to doubly occupy that level. The interaction
between graphene electrons and local level is described by a hybridization
term:
\begin{eqnarray}
H_{V}=V\sum_{\sigma}[f_{\sigma}^{\dagger}\psi_{B,\sigma}(0)+H.c.]\,,
\label{Hvlatt}
\end{eqnarray}
 where $V$ is the hybridization energy, and $\psi_{B,\sigma}(\br)$ 
is one of the components of the spinor defined earlier. 
In (\ref{Hvlatt}) we have assumed that the
adatom hybridizes on top of a carbon atom (in this case, one that is sitting
on sublattice B). This assumption can be easily relaxed and the only difference
is a form factor that takes into account the geometry of the position of the adatom. Since we will be interested in s-wave scattering only, the details about this form factor will be irrelevant in what follows.

The complexity of the problem comes from the $U$ term that describes double occupancy of the adatom. In the absence of this term the problem is quadratic and can be diagonalized exactly \cite{Mahan}. From the point of view of the adatom, graphene behaves as a heat bath that damps the electronic motion in the adatom. As a result, the localized electrons acquire a self-energy, $\Sigma_{ff}(\omega)$,
as a function of frequency $\omega$. The real part of the self-energy is associated with the shift of the local level energy, 
$\epsilon_f = \epsilon_0 - \Re \Sigma_{ff}(\epsilon_0)$, while the imaginary part, $\Im \Sigma_{ff}(\omega)$, gives the decay rate, or broadening, of the adatom level. A straightforward calculation gives: 
\begin{eqnarray}
\Re \Sigma_{ff}(\omega) &=& -V^{2}\frac{\omega}{\Lambda^{2}}\ln\!\left(\frac{\vert\omega^{2}-\Lambda^{2}\vert}{\omega^{2}}\right) \, ,
\nonumber
\\
\Im \Sigma_{ff}(\omega) &=&  - V^{2}\frac{\pi\vert\omega\vert}{\Lambda^{2}}\theta(\Lambda-\vert\omega\vert)\,,
\label{G_bar_R}
\end{eqnarray}
where $\Lambda$ ($\approx \vF/a_0$) is a ultraviolet cut-off. It is immediately clear that the problem at hand is rather different from the case of adatoms in ordinary metals. For one, in ordinary metals the damping of the local level is essentially energy independent and proportional to the DOS at the Fermi level of the metal. In graphene, the DOS vanishes linearly with energy and this is immediately reflected in (\ref{G_bar_R}) where the imaginary part of the self-energy behaves as $|\omega|$ and hence vanishes at the Dirac point (by Kramers-Kronig the real part necessarily has to behave as $\omega \ln(\omega)$ as $\omega \to 0$). This unusual broadening has direct consequences for the LDOS. 

Notice that the LDOS of the adatom (at the position of the adatom) changes from a Dirac delta function, $\rho^{(0)}_{ff,\sigma}(\omega) = \delta(\omega-\epsilon_0)$ to:
\begin{eqnarray}
\rho_{ff,\sigma}(\omega) =  
\frac{1}{\pi}\frac{\Delta\vert\omega\vert Z(\omega)}{[\omega-\epsilon_{0} Z(\omega)]^{2}+(\Delta\vert\omega\vert Z(\omega)+0^{+})^{2}}\,,
\qquad
\label{rhoff}
\end{eqnarray}
where 
\begin{eqnarray}
\Delta=\pi \frac{V^{2}}{\Lambda^{2}}
\end{eqnarray}
is the dimensionless hybridization parameter, and
\begin{eqnarray}
Z^{-1}(\omega)=1+(V^{2}/\Lambda^{2})\ln\left(|\Lambda^{2}-\omega^{2}|/\omega^{2}\right)
\end{eqnarray}
is the quasiparticle residue of the local electrons. One of the most striking features of (\ref{rhoff}) is that, although it is localized at the energy of the level, it has a strong non-lorentzian shape with a substantial tail, something that should be easily measured by STM. Moreover, the quasiparticle residue {\it vanishes} at the Dirac point indicating that the local electron is not a well defined excitation at the Dirac point, being completely merged with the other graphene electrons. These are features that do not show up in ordinary metals. 

The simplest way to include the Coulomb $U$ into the problem is, following Anderson's original paper \cite{anderson}, to perform a mean-field factorization of the interaction term, namely, $n_{\uparrow} n_{\downarrow} \to \langle n_{\uparrow} \rangle n_{\downarrow} + n_{\uparrow} \langle n_{\downarrow} \rangle$, where
\begin{eqnarray}
\langle n_{\sigma} \rangle = \int_{-\Lambda}^{\mu} d\omega \rho_{ff,\sigma}(\omega)
\label{aven}
\end{eqnarray}
is the average adatom occupation that has to be calculated self-consistently. At the mean field level, the only effect of the interaction is the shift the level energy in a spin-dependent way:
\begin{eqnarray}
\epsilon_{\sigma} = \epsilon_f + U \langle n_{-\sigma} \rangle \, ,
\label{reneps} 
\end{eqnarray}
and hence, in order to calculate (\ref{aven}) we can use the non-interacting
result (\ref{rhoff}) by simply replacing the bare level energy, 
$\epsilon_0$, by is renormalized level (\ref{reneps}). The integration in (\ref{aven}) is straightforward and one obtains 
($Z(\epsilon_{\sigma})\equiv Z_{\sigma}$):
\begin{equation}
n_{\sigma}=\frac{Z_{\sigma}^{-1}}{(Z_{\sigma}^{-2}\!+\!\Delta^{2})}\!\!\left[\theta(\frac{\mu}{Z_{\sigma}}\!-\!\epsilon_{\sigma})\!+\! \frac{1}{\pi}\mbox{arctan}\left(\frac{|\mu|\Delta}{\epsilon_{\sigma}\!-\! \frac{\mu}{Z_{\sigma}}}\right)\!+\! \Theta_{\sigma}\right],
\label{MF}
\end{equation}
where
\begin{equation}
\Theta_{\sigma}=Z_{\sigma}\frac{\Delta}{\pi}\ln\!\left[\frac{W_{\sigma}(E_{\sigma})^{\gamma}}{(\epsilon_{\sigma})^{1+\gamma}}\right]-\frac{1}{\pi}\mbox{arctan}\!\left(\frac{\Delta \Lambda}{\Lambda Z_{\sigma}^{-1}+\epsilon_{\sigma}}\right),\label{Theta}\end{equation}
where  $\gamma=\mbox{sign}(\mu)$, 
and 
\begin{eqnarray}
E_{\sigma} & = & \sqrt{(\epsilon_{\sigma}-\mu Z_{\sigma}^{-1})^{2}+\mu^{2}\Delta^{2}}\label{E}\\
W_{\sigma} & = & \sqrt{(\Lambda Z_{\sigma}^{-1}+\epsilon_{\sigma})^{2}+\Lambda^{2}\Delta^{2}}\,.
\label{W}
\end{eqnarray}
The solution of the coupled set of equations (\ref{aven}), (\ref{reneps}), and (\ref{MF}) 
gives the spin population of the local level in the adatom at the mean-field level. Notice that the fact that the energy depends on the spin orientation through (\ref{reneps}) allows for states with different spin occupation and hence with a local moment, $\langle m \rangle$ given by:
\begin{eqnarray}
\langle m \rangle = \langle n_{\uparrow} \rangle - \langle n_{\downarrow} \rangle
\end{eqnarray}
and the states of the adatom can classified as either magnetic, $\langle m \rangle \neq 0$, or non-magnetic, $\langle m \rangle = 0$. One would like to know the boundary line between these two states.

\begin{figure}[btf]
\begin{centering}
\includegraphics[clip,width=8.6cm]{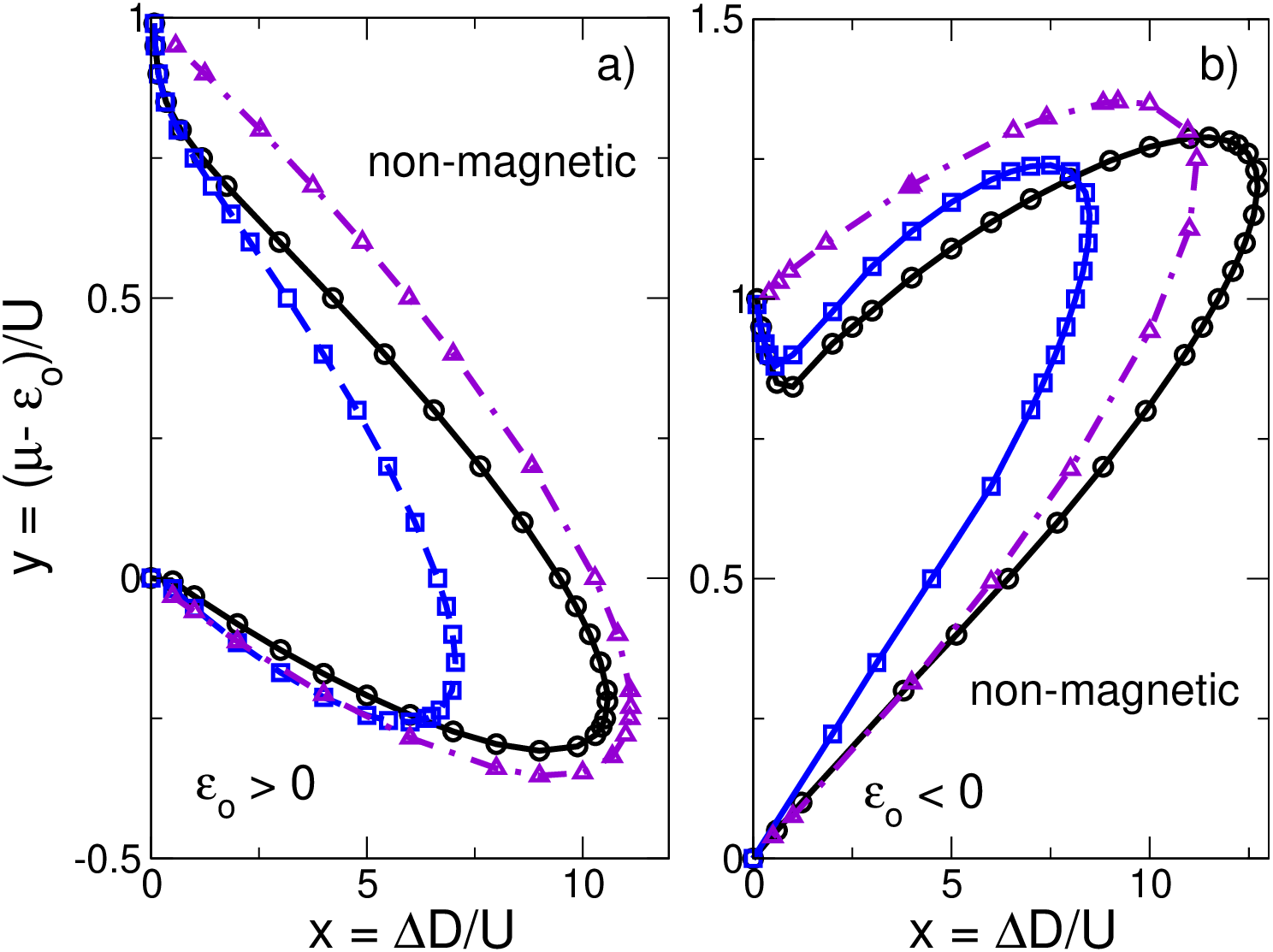} 
\par\end{centering}
\caption{Boundary between magnetic and non-magnetic impurity states in the
scaling variables $x$ and $y$ for $\epsilon_{0}>0$ (a) and $\epsilon_{0}<0$
(b). Circles: $|\epsilon_{0}|/\Lambda=0.029$, $V/\Lambda=0.14$; Squares: $\epsilon_{0}/\Lambda=0.043$
and $V/\Lambda=0.14$; Triangles:$|\epsilon_{0}|/\Lambda=0.029$, $V/\Lambda=0.03$.
The upturn close to $y=1$ and $x\to0$ on panel b) is not visible
in this scale when $V$ is very small (triangles). See details in
the text.  }
\label{Fig_phase} 
\end{figure}

Following Anderson it is convenient to define two dimensionless parameters,
$x=\Lambda\Delta/U$ and $y=(\mu-\epsilon_{0})/U$, that measure the degree of hybridization between adatom and graphene, and the distance in energy of the level from the Fermi energy, respectively. In Fig. \ref{Fig_phase} we show the boundary line between magnetic and non-magnetic states as a function of the parameters of the problem. 

Fig. \ref{Fig_phase} shows that, just like in a normal metal, one needs sufficiently strong Coulomb interactions to generate a magnetic state. Nevertheless there are substantial differences between this diagram and the one obtained in the problem of impurities in ordinary metals \cite{anderson}. 
In an ordinary metal the whole boundary line is essentially scale invariant and depends only on the values of $x$ and $y$. In the case of graphene this is not so and the boundary line also depend on the cut-off $\Lambda$ indicating the sensitivity of the problem to the details in the ultraviolet. This result could be advanced from the form of the self-energy (\ref{G_bar_R}) since it is explicitly dependent on $\Lambda$. Therefore, it would be important to investigate the problem again with the full band of graphene, not only the linearized effective theory (\ref{dirac}). Moreover, clearly the adatom breaks the particle-hole symmetry around the $y=1/2$ line. This comes about because the energy of the localized level and the Fermi energy can be either above or below the Dirac point. On the one hand, then $\epsilon_0>0$ we see that there is a magnetic state even when $y<0$, that is, when $\epsilon_0>\mu$ and the level should be nominally empty. On the other hand, when $\epsilon_0<0$ there is magnetic state when $y>1$, that is, when $\mu > \epsilon_0 +U$ and the level should be nominally doubly occupied. These results contradict our intuition based on the physics of normal metals but once again this can be traced back to the form of the self-energy (\ref{G_bar_R}) which is non-lorentzian with long tails at high energies. These tails allow for formation of local moment even when the position of the energy level is far away from the chemical potential. As a result, it is easier to generate magnetic moments in graphene than in an ordinary metal and one would expect that atoms that are not naturally magnetic on a metal host to become magnetic in graphene.

The application of a potential $V_{g}$ through an electric field
via a back gate\cite{Geim_review} shifts the chemical potential $\mu$
and moves the magnetic state of the impurity in the vertical direction,
that is, the value of $y$ in Fig. \ref{Fig_phase} for a fixed value 
of $x=\Lambda\Delta/U$. In this way, the magnetization of the impurity 
can in principle be turned on and off, depending only on the gate 
voltage applied to graphene. 

\begin{figure}[t]
\begin{centering}
\includegraphics[clip,width=8.7cm]{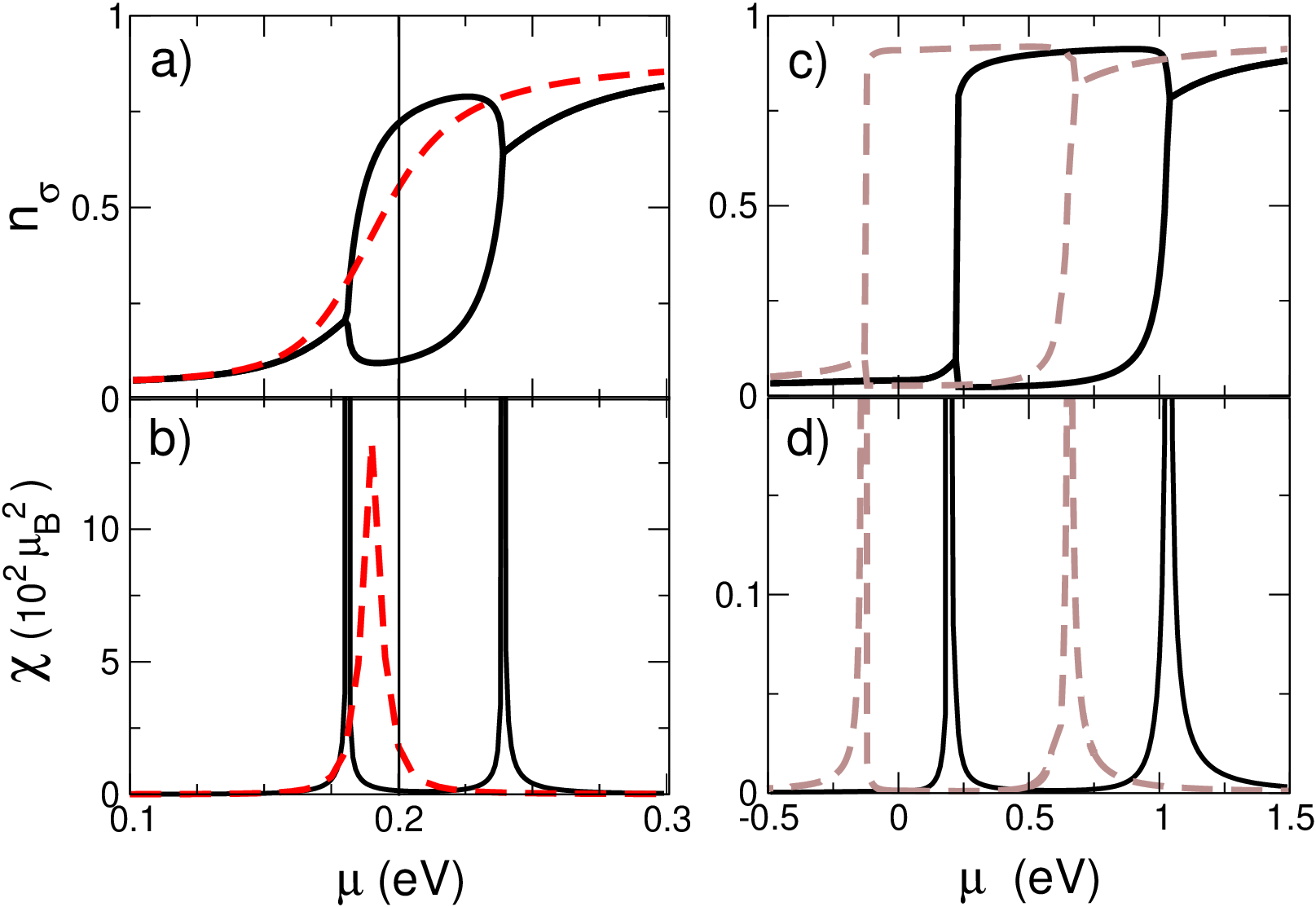} 
\par\end{centering}

\caption{(color on line) Spin polarization, $n_{\uparrow}$,and $n_{\downarrow}$ 
and magnetic susceptibility vs. $\mu$ for |$\epsilon_{0}|/\Lambda \cong0.029$ 
and $V/\Lambda\cong0.14$. Left panels: $x=11$ (dashed curves), and $x=5$
(solid). 
The vertical line marks the position of the level, $\epsilon_{0}=0.2$
eV.  On the right: Comparison between $\epsilon_{0}=0.2$ eV (black
curves) and $\epsilon_{0}=-0.2$ (brown) at $x=0.43$. }
\label{sus} 
\end{figure}

The impurity susceptibility is defined by $\chi=\mu_{B}(n_{\uparrow}-n_{\downarrow})/B$,
where $\mu_{B}$ is the Bohr magneton, and $B$ is the applied magnetic
field. In the presence of a field, the energy of the impurity spin
states changes from (\ref{reneps}) to 
\begin{eqnarray}
\epsilon_{\sigma}=\epsilon_{f}-\sigma\mu_{B}B+Un_{-\sigma} \, .
\end{eqnarray}
In the zero field limit, the magnetic susceptibility of the impurity
$\chi=\mu_{B}\sum_{\sigma}\sigma\left(\mbox{d}n_{\sigma}/\mbox{d}B\right)_{B=0}$
can be calculated straightforwardly from Eq. (\ref{MF}), 
\begin{eqnarray}
\chi=-\mu_{B}^{2}\sum_{\sigma=\uparrow\downarrow}\frac{dn_{\sigma}}{d\epsilon_{\sigma}}\cdot\frac{1-U\frac{dn_{-\sigma}}{d\epsilon_{-\sigma}}}{1-U^{2}\frac{dn_{-\sigma}}{d\epsilon_{-\sigma}}\frac{dn_{\sigma}}{d\epsilon_{\sigma}}}\,,
\label{finalsus}
\end{eqnarray}
which is shown in Fig.\ref{sus}, together with the spin occupation, as a function of chemical potential. The possibility of tuning the magnetism of an adatom with an external electric field is unique to graphene.

The discussion above does not include the Kondo effect, namely, the magnetic screening of the adatom moment by the graphene electrons \cite{SB07,HG07}, and hence it can be thought to be valid for temperatures above the Kondo temperature, $T_K$. The situation here is somewhat reminiscent of the case of magnetic impurities in d-wave superconductors \cite{FFV06} although in graphene the chemical potential is free to vary with the applied gate voltage while in a superconductor the chemical potential is pinned at the superconducting gap. Moreover, the quasiparticle excitations in superconductors have electron-hole character and essentially do not carry charge, while in graphene they do carry charge. More interesting, perhaps, is the fact that the anomalous broadening of the local level puts graphene closer to a mixed valence system, that is, it allows for different oxidation states. In this case, charge fluctuations, which are not usually treated correctly in Kondo-type, spin-only, models, cannot be neglected and one has to deal with the Anderson Hamiltonian directly. It is clear that the calculations described here are only the first step towards the understanding of the magnetic properties of atoms in graphene. With the use of use of the current STM technology one can envision the tantalizing possibility of creating and controlling, at the atomic level, Kondo lattices in the surface of graphene. Since the correlated state of Kondo lattices, the so-called heavy fermion state, is still one of the most intriguing problems in modern condensed matter research \cite{hewson}, graphene presents a possibility for answering many of the questions that cannot be answered in ordinary metal hosts.

\section{Conclusions}

We have shown that the scattering analysis of the Coulomb problem in
graphene has two regimes: undercritical when $g<g_c=1/2$ and supercritical
when $g>g_c$. In these two regimes the behavior of the scattering phase
shifts and the wavefunctions are very different. In the undercritical regime
the physics is more or less conventional, the Coulomb potential breaks
particle-hole symmetry, and leads to an asymmetry in the LDOS close to the adatom. In the supercritical regime resonances show up in the LDOS, mimicking the ``fall to the center'' of an electron with concomitant positron production and, hence,
the charging of the vacuum. It is interesting to note that the transition
between these two regimes is controlled by the charge of the impurity $Z$ and the dielectric properties of the environment, $\epsilon_0$. The supercritical regime may be reached by eliminating dielectric material such as H$_2$O, and working with alkaline atoms such as Ca, that have high oxidation states. 

When electron-electron interactions in the adatom are included via the Anderson impurity model, we have shown that, at the mean-field level, a picture very different from ordinary metal hosts emerges. Due to the anomalous broadening of the local level in graphene, magnetic states that are nominally not allowed can become magnetic. This opens the possibility for the magnetization of adatoms that are not usually magnetic in normal metals. Since graphene is an excellent conductor and magnetism can theoretically be achieved, one can envision using graphene for spintronics applications \cite{wolf} where graphene works as the conductor of charge, and the adatoms work as magnetic memory. In this case, we would have the logical and memory materials embedded in the same matrix, but more important, and unlike diluted magnetic semiconductors \cite{allan}, the position of the adatoms can be controlled with atomic precision with a STM.

It is clear from the previous discussions that, no matter how 
we look at the Coulomb problem in graphene, either from the perspective
of scattering theory, or magnetism, it has many unusual features that
do not appear in conventional metals. Properties such as
supercritical behavior and ``fall to the center'' physics, and magnetization
in nominally forbidden regions of the magnetic diagram, are not
intuitive and emerge in ordinary matter only when those are subject to extreme 
conditions such as enormous electric fields. Nevertheless, in graphene 
such extraordinary
properties can be measured with standard condensed matter experimental 
probes such as scanning tunneling microscopy. Moreover, such unusual
properties may be helpful in order to generate strongly-correlated states
by chemically modifying graphene, a field that is still in its infancy.

AHCN acknowledges the partial support of the U.S. Department of Energy under
grant DE-FG02-08ER46512. NMRP acknowledge the financial support from POCI 2010 via project PTDC/FIS/64404/2006. VMP is supported by FCT via SFRH/BPD/27182/2006 and PTDC/FIS/64404/2006.


\bibliography{graphene_coulomb}

\begin{thebibliography}{44}
\bibitem[{\citenamefont{{The critical value of $Z$ has also corrections due to
  electron-electron interactions.}}()}]{note}
\bibinfo{author}{\bibnamefont{{The critical value of $Z$ has also corrections
  due to electron-electron interactions.}}}
\end{thebibliography}

\end{document}